\documentclass[twocolumn,prb]{revtex4}
\usepackage{graphicx}
\usepackage{dcolumn}
\usepackage{bm}
\usepackage{amsmath}

\begin{document}

\preprint{}
\title{Tuning odd triplet superconductivity by spin pumping}
\author{Takehito Yokoyama$^{1}$ and Yaroslav Tserkovnyak$^{2}$}
\affiliation{$^1$Department of Applied Physics, University of Tokyo, Tokyo 113-8656, Japan \\
$^2$ Department of Physics and Astronomy, University of California, Los Angeles, California 90095, USA}
\date{\today}

\begin{abstract}
We study proximity effect in diffusive ferromagnet/normal-metal/superconductor junction with precessing magnetization of the ferromagnet. We find that the odd-frequency pairing induced in the normal metal is modified by spin pumping from the ferromagnet and hence can be tuned by changing the precessional frequency. 
At the frequency corresponding to twice the superconducting gap, the odd-frequency pairing is strongly enhanced.  We find a crossover from the even- to the odd-frequency superconductivity in the normal metal by tuning the precessional frequency. This gives a clearcut signature of the odd-frequency superconductivity observable by scanning tunneling microscopy. According to the pairing symmetries in the normal metal, we find  a crossover from the gap to the peak structure in the tunneling conductance between the normal metal and a scanning tunneling microscope tip. 
\end{abstract}

\pacs{PACS numbers:75.75.+a, 73.20.-r, 75.50.Xx, 75.70.Cn}
\maketitle



%

%




\section{Introduction}
Generally, superconducting correlations can be even  
or odd in frequency, depending on their symmetry with respect to the  
time axis. In accordance with the fermionic statistics, even-frequency 
superconductors are characterized by the spin-singlet even-parity
or spin-triplet odd-parity pairing states, while odd-frequency
superconductors are grouped into the spin-singlet odd-parity
or spin-triplet even-parity pairing states.

The possibility of the odd-frequency pairing state in a 
uniform system was discussed in the literature, \cite{Berezinskii,Balatsky,Coleman,Vojta,Fuseya} although its realization in bulk materials is still controversial. The odd-frequency pairing state has recently been predicted in inhomogeneous superconducting systems. \cite{Efetov1,Kadigro,Tanaka,Tanaka3,EschrigLTP,eschrig,Yokoyama,Linder}
In diffusive ferromaget/superconductor junctions, odd-frequency pairings emerge due to the broken symmetry in spin space.\cite{Efetov1,Kadigro} 
In the diffusive ferromagnet, only $s$-wave symmetry of the pair amplitude is allowed due to impurity scattering and hence triplet pairing inevitably belongs to the odd-frequency superconductivity, which is also called \textit{odd triplet superconductivity}.

Recently, the interplay between ferromagnetism and superconductivity in ferromagnet/normal-metal/superconductor (F/N/S) junctions has been studied using spin-active boundary condition.\cite{hh} Another recent progress is the study of the Josephson effect in S/F/S junctions, where it has been established that the ferromagnetic spin dynamics play a crucial role.\cite{maekawa,Houzet,Konschelle} The F/N/S proximity effect in the presence of spin dynamics, however, still remains to be fully understood. Moreover, although the long-range proximity effect by the generation of the odd-frequency triplet pairing has been observed in recent experiments,\cite{keizer,sosnin} no experiment has succeeded in controlling the magnitude of the odd-frequency pairing to this date. 

In this paper, we propose an experimental setup which makes it possible to tune the magnitude of the odd-frequency pairing. 
We study the proximity effect in diffusive F/N/S junctions with precessing magnetization of the F layer. We find that the odd-frequency pairing induced in the N layer is modified by spin pumping from the F and therefore can be tuned by changing the precessional frequency. 
At the resonance frequency (i.e., twice the superconducting gap), the odd-frequency pairing is strongly enhanced.  We find a crossover from the even- to odd-frequency superconductivity in the N layer by changing the precessional frequency. This is reflected in the tunneling conductance between the normal metal and the scanning tunneling microscope (STM) tip as a crossover from the gap to the peak structure. 

\section{Formulation}
\begin{figure}[htb]
\begin{center}
\scalebox{0.4}{\includegraphics[width=19.0cm,clip]{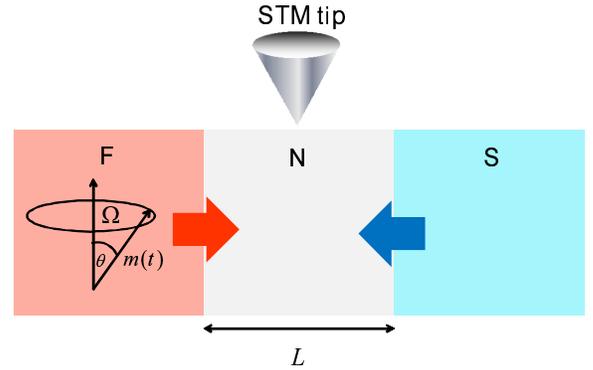}}
\end{center}
\caption{(Color online) Schematic of the model for a ferromagnet/normal-metal/superconductor (F/N/S) junction with precessing magnetization of the F layer. A fictitious exchange field is dynamically induced in the N layer, which interplays with the superconductivity induced by the proximity effect. The electronic structure in the N layer can be studied by scanning tunneling microscopy (STM).}
\label{f1}
\end{figure}

We consider a diffusive F/N/S junction with precessing magnetization of the F as shown in Fig.~\ref{f1}. The F/N interface is located at $x=0$ while the N/S interface is at $x=L$. The spin relaxation in the junction is assumed to be weak. Then, a spin density is pumped into the N from F, \cite{Tserkovnyak} while the superconductivity is  induced in the N by the proximity effect. Therefore, we can study the interplay between ferromagnetism and superconductivity in the N layer.\cite{hh}

Let us take the time-dependent exchange field in the F to be directed along
\begin{eqnarray}
\mathbf{m}(t)= (\sin \theta \cos \Omega t,\sin \theta \sin \Omega t,\cos \theta )\,.
\end{eqnarray}
Here, $\Omega$ is the precessional frequency around $z$ axis and $\theta$ is a constant tilt angle. In the rotating frame, such precession can be viewed as a difference between spin-resolved chemical potentials along the $z$ axis, and the noncollinear effective exchange fields in N and F can generate a long-range proximity effect.\cite{Braude,houzet2,Houzet} To investigate the proximity effect, we use the unitary transformation:
\begin{eqnarray}
\hat g(t,t') \to U^\dag(t)\hat g(t,t')U(t')\,,
\end{eqnarray}
where $U(t) = \exp \left(-i\Omega t\sigma_3/2\right)$ transforms from the laboratory into the spin-rotating frame. Then, the problem reduces to the stationary one. \cite{Houzet}
In fact, we have $U^\dag(t)\mathbf{m}(t) \cdot \bm{\sigma} U(t) = \mathbf{m}(0) \cdot \bm{\sigma}  \equiv \mathbf{m} \cdot \bm{\sigma}$. Here, $\hat g$ is the retarded component of the quassiclassical Green's function, which is 4$\times$4 matrix in spin$\otimes$Nambu space. We parameterize $\hat g$ as
\begin{equation}
\hat g = \tau _3  \otimes (g + {\mathbf{g}} \cdot {\bm{\sigma }}) + \tau _1 \otimes (f_s  + {\mathbf{f}}_t  \cdot {\bm{\sigma }})\,,
\label{par}
\end{equation}
where $g$ and $f_s$ are  scalars, and $\mathbf{g}$ and $\mathbf{f}_t$ are three-dimensional vectors.\cite{champel} $\sigma _i$ and $\tau _i$ $(i = 0,1,2,3)$ are the unit and Pauli matrices in the spin and Nambu spaces, respectively. $\bm{\sigma}=(\sigma_1,\sigma_2,\sigma_3)$ is the vector of Pauli matrices.

The Usadel equation in the N in this rotating frame has the form \cite{usadel,ivanov_prb_06,Houzet}
\begin{eqnarray}
D\boldsymbol{\nabla} (\hat g \boldsymbol{\nabla} \hat g) + \left[ {i\varepsilon \tau _3  + i(\Omega /2)\sigma _3 \otimes \tau _3 ,\hat g} \right] = 0\,,
\label{usadel}
\end{eqnarray}
where $D$ is the diffusion constant and $\varepsilon$ is the quasiparticle energy measured from the Fermi level. See Appendix~\ref{AUE} for derivation. To take into account the magnetic proximity effect, we consider a low-transparency spin-active interface at the F/N contact, with the boundary condition on the N side given by\cite{hh,Cottet}
\begin{eqnarray}
2\gamma _B \xi \hat g \partial_x \hat g = \left[ {\tau _3  + i\gamma _\phi(\mathbf{m} \cdot \bm{\sigma}  ) \otimes \tau _3 ,\hat g} \right]\,.
\end{eqnarray}
Here, $\gamma _B=R_b L/R_d \xi$ and  $\gamma _\phi   = G_\phi/G_T$, in terms of the interfacial resistance parameter $R_b$, the diffusive resistance of the N $R_d$, the N coherence length $\xi$, the imaginary part of the mixing conductance\cite{Brataas} $G_\phi$ and the interfacial tunneling conductance $G_T\sim1/R_b$. The magnetic proximity effect is governed by $\gamma _\phi$.\cite{hh} We will for simplicity disregard the real part of the mixing conductance and the spin dependence of the F/N conductance, which should not affect our findings qualitatively. We make use of the well-known Kupriyanov-Lukichev boundary conditions\cite{kupluk} for the N/S interface:
\begin{eqnarray}
-2\gamma _B \xi \hat g \partial_x \hat g = \left[ {\hat g_S ,\hat g} \right]\,,
\end{eqnarray}
with the bulk Green's functions $\hat g_S$. For simplicity, we assume the same $\gamma_B $ parameter at both interfaces.

Focusing on the tunneling regime with weak superconducting correlations in N, we linearize the Usadel equation with respect to the anomalous Green's function in the N. The linearized Usadel equation reads
\begin{eqnarray}
D\partial_x^2 f_s  + 2i \varepsilon f_s  - 2i\mathbf{f}_t  \cdot \mathbf{h} &=& 0\,,\nonumber\\
D\partial_x^2 \mathbf{f}_t  + 2i \varepsilon \mathbf{f}_t  - 2if_s \mathbf{h} &=& 0\,,
\end{eqnarray}
with $\mathbf{h} = -\Omega\mathbf{z}/2$ and $\mathbf{z}=(0,0,1)$.
The general solution in the N is thus
\begin{eqnarray}
\left( {\begin{array}{*{20}c}
   {f_s }  \\
   {\mathbf{f}_t }  \\
\end{array}} \right) \equiv \left( {\begin{array}{*{20}c}
   {f_s }  \\
   {f_1 }  \\
   0  \\
   {f_3 }  \\
\end{array}} \right) &=& (Ae^{ik_ +  x}  + A'e^{ - ik_ +  x} )\left( {\begin{array}{*{20}c}
   1  \\
   0  \\
   0  \\
   { - 1}  \\
\end{array}} \right) \nonumber \\ &+& (Be^{ik_ -  x}  + B'e^{ - ik_ -  x} )\left( {\begin{array}{*{20}c}
   1  \\
   0  \\
   0  \\
   1  \\
\end{array}} \right) \nonumber\\ &+& (Ce^{ik_0 x}  + C'e^{ - ik_0 x} )\left( {\begin{array}{*{20}c}
   0  \\
   1  \\
   0  \\
   0  \\
\end{array}} \right)\,,
\label{solution}
\end{eqnarray}
with $k_ \pm   = \sqrt {2i( - \varepsilon  \pm \Omega /2)/D}$ and $k_0  = \sqrt { - 2i\varepsilon /D}$. The coefficients $A, A', B, B', C$ and $C'$ are determined by the boundary conditions. 
Here, $f_1$ is the transverse triplet component, which is long ranged at low energies, and $f_3$ is the longitudinal triplet component, which is short ranged, similarly to the singlet component $f_s$.\cite{Efetov1} It follows from the boundary conditions that the second component of $\mathbf{f}_t$ is zero.

Linearizing the boundary conditions at the N/F interface, we have
\begin{eqnarray}
 \gamma _B \xi \partial _x f_s  &=& f_s  + i \gamma_\phi \mathbf{f}_t  \cdot \mathbf{m}\,, \nonumber \\ 
 \gamma _B \xi \partial _x \mathbf{f}_t  &=& \mathbf{f}_t  + i \gamma_\phi f_s \mathbf{m}\,.
 \label{bc2}
\end{eqnarray}
These boundary conditions show that the presence of $f_s$ generates the triplet components with $\mathbf{f}_t \parallel \mathbf{m}$ as long as $\gamma _\phi   \ne 0$. Similarly, the linearized boundary conditions at the N/S interface have the form\begin{eqnarray}
 \gamma _B \xi \partial _x f_s  + g_S^0 f_s  + g_S^3 f_3  &=& f_S^0 \,,\nonumber \\ 
\gamma _B \xi \partial _x {\bf{f}}_t  + g_S^0 {\bf{f}}_{t}  + g_S^3 f_s {\bf{z}} &=& f_S^3 {\bf{z}}\,.
\label{bc4}
\end{eqnarray}
Here, the bulk Green's functions in the S, $g_S$ and $f_S$, are given by $g_S  = \left( {g_S^0 \sigma _0  + g_S^3 \sigma _3 } \right) \otimes \tau _3$ and $ f_S  = \left( {f_S^0 \sigma _0  + f_S^3 \sigma _3 } \right) \otimes \tau _1$, where (in a convenient gauge)
\begin{eqnarray}
g_S^0 &=& \frac{1}{2}\left( {\frac{{ - i(\varepsilon + \Omega /2)}}{{\sqrt {\Delta ^2  - (\varepsilon + \Omega /2)^2 } }} + \frac{{ - i(\varepsilon - \Omega /2)}}{{\sqrt {\Delta ^2  - (\varepsilon - \Omega /2)^2 } }}} \right)\,,\nonumber \\ 
 g_S^3  &=& \frac{1}{2}\left( {\frac{{ - i(\varepsilon + \Omega /2)}}{{\sqrt {\Delta ^2  - (\varepsilon + \Omega /2)^2 } }} - \frac{{ - i(\varepsilon - \Omega /2)}}{{\sqrt {\Delta ^2  - (\varepsilon - \Omega /2)^2 } }}} \right)\,,\nonumber \\ 
f_S^0  &=& \frac{1}{2}\left( {\frac{\Delta }{{\sqrt {\Delta ^2  - (\varepsilon + \Omega /2)^2 } }} + \frac{\Delta }{{\sqrt {\Delta ^2  - (\varepsilon - \Omega /2)^2 } }}} \right)\,,\nonumber \\ 
 f_S^3  &=& \frac{1}{2}\left( {\frac{\Delta }{{\sqrt {\Delta ^2  - (\varepsilon + \Omega /2)^2 } }} - \frac{\Delta }{{\sqrt {\Delta ^2  - (\varepsilon - \Omega /2)^2 } }}} \right)\,.\nonumber\\
\label{ggff}
\end{eqnarray}
After the unilluminating manipulations to solve Eqs.~(\ref{solution})-(\ref{bc4}), we obtain the desired solution of the Usadel equation. The explicit form of the solution is rather complicated and is given in Appendix~\ref{SUE}.


In general, the tilt angle $\theta$ is determined by solving the Landau-Lifshitz-Gilbert equation:
\begin{eqnarray}
\frac{{d\mathbf{m}(t)}}{{dt}} =  - \gamma \mathbf{m}(t) \times \mathbf{H}_{\rm eff}(t)  + \alpha \mathbf{m}(t) \times \frac{{d\mathbf{m}(t)}}{{dt}}\,,
\end{eqnarray}
where $\gamma$ is the gyromagnetic ratio, $\mathbf{H}_{\rm eff}$ is the effective magnetic field and  $\alpha$ is the Gilbert damping constant. 
Driving the ferromagnet by an applied transverse rotating magnetic field $h_{\rm rf}$, in the presence of an effective static dc field $H_{\rm dc}$, we find a small transverse magnetic field in the rotating frame of reference. This is disregarded in our analysis, assuming $\alpha\ll1$, so that $\theta \sim h_{\rm rf}/\alpha H_{\rm dc}\gg h_{\rm rf}/H_{\rm dc}$ at the ferromagnetic resonance, which is usually the case in realistic ferromagnets. We note that the spin-pumping effect is enhanced close to the ferromagnetic resonance, which should in practice be the optimal driving regime for the magnetic dynamics.
Note that the tilt angle of the rotating magnetization $\theta$ may in general be time dependent. In this paper, we focus on the regime when
$d\ln\theta/dt\ll(\Omega,\theta^2/\tau_i)$, so that the appropriate nonequilibrium spin state is fully developed before $\theta$ changes appreciably. Here, $1/\tau_i$ is an effective spin-injection rate proportional to the mixing conductance at the F/N interface.\cite{Brataas2} In addition, the spin-relaxation rate is assumed to satisfy $1/\tau_{\rm sf}\ll\Omega$, so that the spin memory is preserved during a cycle of precession and $1/\tau_{\rm sf}\ll\theta^2/\tau_i$, so that the developed spin accumulation does not decay. We will, therefore, fix  $\theta$ and disregard spin relaxation in the junction in the following.

Next, let us consider tunneling current between the N and the STM tip. See Fig.~\ref{f1}. Below, we consider two cases: spin relaxation in the STM tip is weak or strong, on the scale of its characteristic spin-injection rate. Weak (strong) spin relaxation effectively leads to a state of equilibrium in the STM tip, in the rotating (lab) frame of reference. To realize the situation that the state of equilibrium is effectively reached in the rotating frame of reference, the STM tip should be spin-polarized by the fictitious field, which would in practice require either connecting the  precessing ferromagnet directly to the tip or using magnetized tip which itself is precessing in synchronization.  Experimentally, this may be more difficult to realize than the strong spin-relaxation regime in the STM tip. (Note, however, that a spin-resolved STM technique has been recently developed.\cite{Meier})
One may furthermore think that the limit when $\Omega \sim \Delta$ is problematic for the superconductor, since the gap should then be strongly suppressed. However, assuming some spin relaxation in S, the out-of-equilibrium pumped spins do not cause significant Cooper-pair depairing in the superconductor, while the effective spin relaxation in N is weak, so that the pumped spin distribution is unaffected in N.

Now, let us first consider the case of the STM tip with weak spin relaxation.
We define the 8$\times$8 matrix Green's functions in STM and N, $\check{g}_{STM}$ and $\check{g}_N $, as
\begin{eqnarray}
\check{g}_{STM}  = \left( {\begin{array}{*{20}c}
   {\tau _3 } & {\hat g_{STM}^K }  \\
   0 & { - \tau _3 }  \\
\end{array}} \right)\,,\,\,\, \check{g}_N  = \left( {\begin{array}{*{20}c}
   {\hat g_N^R } & {\hat g_N^K }  \\
   0 & {\hat g_N^A }  \\
\end{array}} \right)\,.
\end{eqnarray}
Here, $R,K$ and $A$ represent the retarded, Keldysh and advanced components, respectively. 
The current between the N and the STM tip is given by
\begin{eqnarray}
I \sim \int {d\varepsilon {\rm{Tr}}} \left( {\tau _3 \left[ {\check{g}_{STM} ,\check{g}_N } \right]^K } \right)\,,
\end{eqnarray}
where
\begin{widetext}
\begin{eqnarray}
\left[ {\check{g}_{STM} ,\check{g}_N } \right]^K  = \hat g_{STM}^K \hat g_N^A  - \hat g_N^R \hat g_{STM}^K  + \tau _3 \hat g_N^K  + \hat g_N^K \tau _3\,.
\end{eqnarray}
Let us apply a bias voltage $V$ to the STM tip, so that we have
\begin{eqnarray}
\hat g_{STM}^K  &=& \tau _3 \left( {f_{STM}^0  + f_{STM}^3 \tau _3 } \right) + \left( {f_{STM}^0  + f_{STM}^3 \tau _3 } \right)\tau _3  = 2f_{STM}^3  + 2f_{STM}^0 \tau _3\, , \hfill\nonumber \\
  f_{STM,\sigma }^{0(3)} &=& \left[ {\tanh \left\{ {(\varepsilon  + eV + \sigma \Omega /2)/2T} \right\} + ( - )\tanh \left\{ {(\varepsilon  - eV - \sigma \Omega /2)/2T} \right\}} \right]/2\,, \label{f}
\end{eqnarray}
with $ \sigma  =  \pm $ for spins up and down, and $\hat g_N^K  = \tanh (\varepsilon /2T)\left( {\hat g_N^R  - \hat g_N^A } \right)$.
Therefore, we obtain
\begin{eqnarray}
{\text{Tr}}\left( {\tau _3 \left[ {g_{STM} ,g_N } \right]^K } \right) = \sum\limits_\sigma  {2f_{STM,\sigma }^3 \left( {g_{N,\sigma }^A  - g_{N,\sigma }^R } \right)}  =  - \sum\limits_\sigma  {4f_{STM,\sigma }^3 \operatorname{Re} g_{N,\sigma }^R }\,,
\end{eqnarray}
and, hence, at zero temperature: 
\begin{eqnarray}
dI/dV \sim \sum\limits_\sigma \left[ {N_\sigma  \left( { - eV - \sigma \Omega /2} \right) + N_\sigma  \left( {eV + \sigma \Omega /2} \right)}\right]\,. \label{cond}
\end{eqnarray}
\end{widetext}
Here, $N_\sigma ={\rm Re}g^R_{N,\sigma}$ denotes the density of states for spin $\sigma $ in the N. 
To calculate the spin-resolved density of states $N_\sigma$, we can use the relation ${\mathbf{g}} =  - f_s {\mathbf{f}}_t $, which is obtained from the normalization condition $\hat{g}^2=1$ applied to the parametrization (\ref{par}).


In the case of the STM tip with weak spin relaxation, the state in the STM is equilibrated in the rotating frame of reference, and we obtain the relevant results by setting $\Omega=0$ in Eqs.~(\ref{f}) and (\ref{cond}). The tunneling conductance then coincides exactly with the density of states. 

\section{Results}
In the numerical calculations, we introduce a small imaginary part of the energy, $\varepsilon  \to \varepsilon  + i\delta$, to regularize singularities. Physically, $\delta$ captures an effective depairing rate for Cooper pairs. 
We choose the following parameters: $\gamma _B=100$, $\theta=\pi/12$, $L/\xi=10$, $\Delta/E_{\rm Th}=0.1$, and $\delta/E_{\rm Th}=0.01$. Here, $E_{\rm Th}=D/L^2$ is the Thouless energy. Typically, for $L/\xi=10$, $E_{\rm Th}\sim 0.01$~meV.

\subsection{Weak spin relaxation in the STM tip}
\label{WSR}

We start by considering the case of weak spin relaxation in the STM tip. 
We first discuss some limiting cases and give qualitative picture of our main results. Below, we focus on the low-energy limit, i.e., $\varepsilon\to0$, where the superconductivity is manifested most strongly. 
At $\varepsilon =0$, the bulk Green's functions in the S are given by
\begin{eqnarray}
 g_S^0  &=& 0\,,\,\,\,g_S^3  = \frac{{ - i\Omega /2}}{{\sqrt {\Delta ^2  - (\Omega /2)^2 } }}, \\ 
 f_S^0  &=& \frac{\Delta }{{\sqrt {\Delta ^2  - (\Omega /2)^2 } }}\,,\,\,\,f_S^3  = 0\,,
\end{eqnarray}
for $\Omega  < 2\Delta$ and
\begin{eqnarray}
 g_S^0  &=& \frac{{\Omega /2}}{{\sqrt {(\Omega /2)^2  - \Delta ^2 } }}\,,\,\,\,g_S^3  = 0, \\ 
 f_S^0  &=& 0\,,\,\,\,f_S^3  = \frac{{i\Delta }}{{\sqrt {(\Omega /2)^2  - \Delta ^2 } }}\,,
\end{eqnarray}
 for $\Omega  > 2\Delta$. In the limit of $\Omega  \to 2\Delta$, these Green's functions diverge and the proximity effect is resonantly enhanced as seen from the boundary conditions at the N/S interfaces. Taking the limit $\gamma _\phi   \to \infty$ for the boundary condition at the F/N interface, we have $f_s  \to 0$ and $f_1 \sin \theta  + f_3 \cos \theta  \to 0$. The singlet component vanishes while the triplet components can remain finite for  $\theta  \ne 0, \pi /2$ in this limit.  
Thus, we can expect that  by tuning $\Omega$ close to $ 2\Delta$ one can control the magnitude of the odd-frequency superconductivity for sufficiently large  $\gamma _\phi$.


\begin{figure}[htb]
\begin{center}
\scalebox{0.4}{\includegraphics[width=22.0cm,clip]{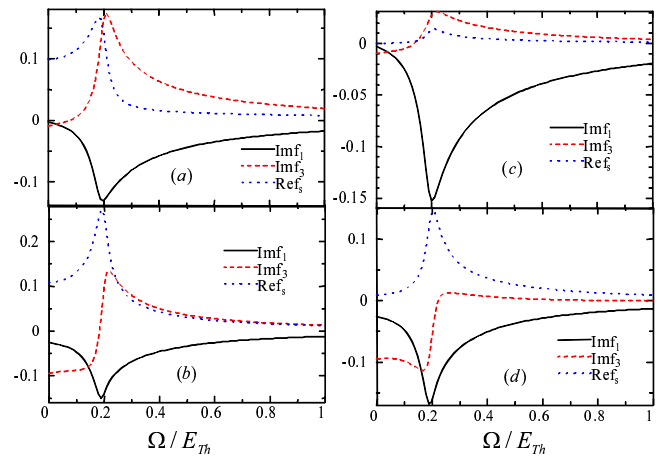}}
\end{center}
\caption{(Color online) Anomalous Green's functions as a function of $\Omega /E_{\rm Th}$  at the N/S interface ($x=L$) with (a) $\gamma _\phi=100$ and (b) $\gamma _\phi=10$ and at the F/N interface ($x=0$) with (c) $\gamma _\phi=100$ and (d) $\gamma _\phi=10$. Here, we set $\varepsilon=0$. $f_1$ and $f_3$ are the odd-frequency and $f_s$ is the even-frequency pairing amplitudes. } \label{f2}
\end{figure}

\begin{figure}[htb]
\begin{center}
\scalebox{0.4}{\includegraphics[width=17.0cm,clip]{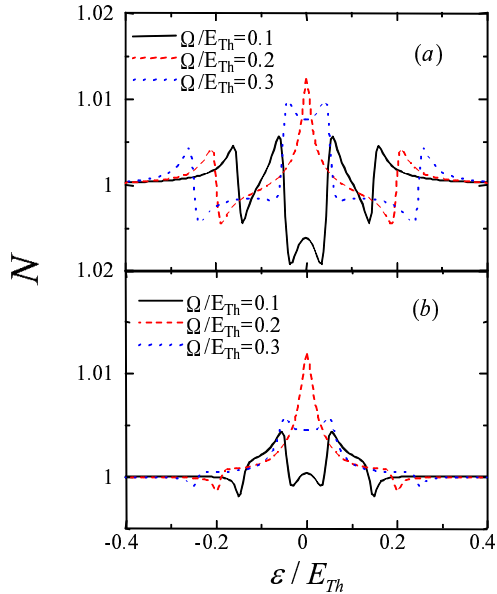}}
\end{center}
\caption{(color online) The density of states normalized by its normal state value as a function of $\varepsilon /E_{\rm Th} $ with $\gamma _\phi=100$ at (a) $x=L$ and (b) $x=0$. } \label{f3}
\end{figure}

Next, let us plot the anomalous Green's functions using Eq.~(\ref{solution}).
Figure~\ref{f2} shows anomalous Green's functions at the N/S interface ($x=L$) as a function of $\Omega /E_{\rm Th} $  with (a) $\gamma _\phi=100$ and (b) $\gamma _\phi=10$. Note that at $\varepsilon =0$, $f_1$ and $f_3$ are purely imaginary while $f_s$ is  a real number. We find that a resonant peak appears at $\Omega=2\Delta$. As seen in Fig.~\ref{f2}(a),  for $\Omega<2\Delta$, the singlet component dominates, while, for $\Omega > 2\Delta$, the triplet component dominates. This can be understood by the fact that for $\Omega  \to 2\Delta  - 0^+$, $f_S^0$ diverges while for  $\Omega  \to 2\Delta  + 0^+$, $f_S^3$ diverges. 
\textit{Thus, one can control a crossover from the even- to the odd-frequency superconductivity by changing $\Omega$, which is tunable by the external magnetic field.} When $\gamma _\phi$ is reduced, the singlet component is enhanced as shown in Fig.~\ref{f2}(b). We also show the anomalous Green's functions at the F/N interface ($x=0$) as a function of $\Omega /E_{\rm Th} $ in Fig.~\ref{f2}(c) for $\gamma _\phi=100$ and Fig.~\ref{f2}(d) for $\gamma _\phi=10$. A peak also appears at $\Omega=2\Delta$. Compared to the results at $x=L$, the long-range triplet component $f_1$ has a large magnitude, which is controllable by tuning the frequency $\Omega$.

To date, a hallmark of the odd-frequency pairing has been considered to be the long-ranged proximity effect in the presence of magnetism.\cite{Efetov1} However, another aspect of this pairing has been recently appreciated: The density of states in the presence of the odd-frequency pairing is enhanced, acquiring a zero-energy peak within the gap structure.\cite{Tanaka,Asano,Yokoyama2,Braude} Using general relations for the conjugate Green's functions,\cite{Eschrig} we have that $ \tilde f_s (\varepsilon ) = f_s^* ( - \varepsilon )$ and $\tilde {\mathbf{f}}_t (\varepsilon ) =  - \mathbf{f}_t^* ( - \varepsilon )$. Hence, we easily obtain $g^2  = 1 - \left| {f_s(0)} \right|^2  + \left| {\mathbf{f}_t(0)} \right|^2$ at $\varepsilon=0$, from the standard normalization condition $\hat{g}^2=1$. Therefore, the density of states,  which is given by ${\rm Re}\,g$, is enhanced by the generation of odd-frequency pairing ($\mathbf{f}_t$) and suppressed by the presence of the even-frequency pairing ($f_s$) at $\varepsilon=0$. 

The density of states normalized by its normal state value, $N = {\mathop{\rm Re}\nolimits} \sqrt {1 - (f_s^2  + f_1^2  + f_3^2 )}$, as a function of $\varepsilon /E_{\rm Th} $ is shown in Fig.~\ref{f3} setting $\gamma _\phi=100$ at (a) $x=L$ and (b) $x=0$. We find a crossover from the gap to the peak structure in the density of states with increasing $\Omega /E_{\rm Th}$. This reflects a crossover from the even- to the odd-frequency superconductivity in the N, upon changing the precessional frequency. Note that, as shown above, the density of states exactly coincides with the tunneling conductance between the N and the STM tip, in the weak spin-relaxation regime of the STM tip. 

\subsection{Strong spin relaxation in the STM tip}

\begin{figure}[htb]
\begin{center}
\scalebox{0.4}{\includegraphics[width=17.0cm,clip]{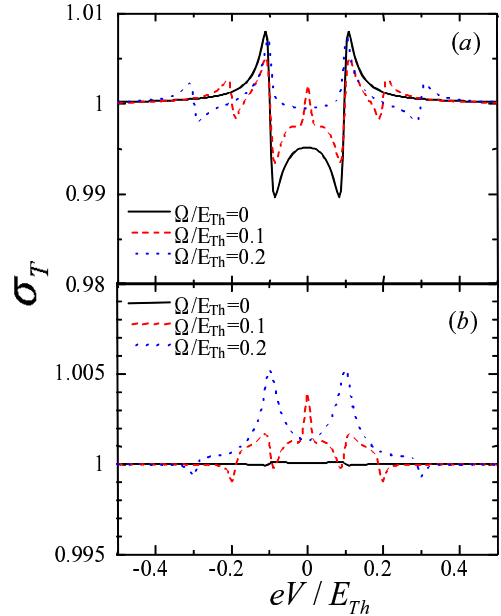}}
\end{center}
\caption{(color online) The normalized tunneling conductance as a function of $eV/E_{\rm Th} $ with $\gamma _\phi=100$ at (a) $x=L$ and (b) $x=0$. } \label{f4}
\end{figure}

Here, we discuss the more realistic case with strong spin relaxation in the STM tip, so that the tip is equilibrated in the lab frame. 
Let us define $\sigma_T$ as $dI/dV$ normalized by its normal state value. 
 Figure~\ref{f4} shows  $\sigma_T$ as a function of $eV/E_{\rm Th} $ for $\gamma _\phi=100$ at (a) $x=L$ and (b) $x=0$. 
We see the gap structure at $\Omega=0$. For larger $\Omega$, the zero-bias conductance grows, and a sharp zero-bias peak emerges at $\Omega/E_{\rm Th}=0.1$. This can be understood as follows: 
Let us consider  the term $N_{\uparrow}  (eV + \Omega /2)$  in Eq.~(\ref{cond}), for instance. Then, the argument $\varepsilon - \Omega /2$ appearing in Eqs.~(\ref{usadel}) and (\ref{ggff})  is effectively replaced by $eV $ while $\varepsilon + \Omega /2$ is replaced by $eV + \Omega$. Similar replacements should be made for other terms in Eq.~(\ref{cond}). Thus, coherence peaks will appear at $ eV=\Omega \pm \Delta$. Therefore, when $\Omega = \Delta$, a resonant peak appears at the zero bias. This can be also attributed to the emergence of the odd-frequency pairing, as shown in Sec.~\ref{WSR}. Therefore, we find that even in the case of the strong spin relaxation in the STM, we can also find the evidence of the odd triplet superconductivity (zero-bias peak) tuned by the precession frequency. 



\section{Conclusions}

We have studied the proximity effect in diffusive F/N/S junctions with precessing magnetization of the F layer. We found that a tunable odd-frequency pairing in the N is governed by spin correlations in the N that are induced by the ferromagnetic precession. At the resonance frequency, $\Omega\to2\Delta$, the odd-frequency pairing is strongly enhanced.  We unveiled a crossover from the even- to the odd-frequency superconductivity in the N by tuning the precessional frequency. 
 This is manifested as the crossover from the gap to the peak structure in the tunneling conductance between the N and the STM tip, where we investigated two regimes of the weak and strong spin relaxation in the STM tip.

Our results can be experimentally accessible and provide a novel way to amplify the odd triplet superconductivity using the magnetization precession in the F/N/S junctions. These characteristics should be observable by a scanning tunneling microscope or other tunneling experiments and, hence, may serve as a smoking gun to reveal the odd-frequency superconductivity.

Recently, the F/N/S trilayer junctions have been fabricated to study behavior of the superconducting phase transition.\cite{Yamazaki,Kim}
Also, a ferromagnetic resonance experiment has been performed in an Nb/permalloy proximity system.\cite{aarts}
In the light of these advances, it appears realistic to verify our predictions experimentally by using, e.g., permalloy/Cu/Nb junctions. 
The predicted resonance frequency corresponds to the $\sim$MHz range and is easily achievable by the present-day experimental technique. It should be remarked, however, that the required condition for the spin-flip relaxation rate, $1/\tau_{\rm sf}\ll\Omega$, is rather stringent in this low-frequency limit, requiring low temperatures and very clean normal interlayer.

\acknowledgements

We are grateful to A.~Brataas, Ya.~V. Fominov, M.~Houzet, J.~Linder and J.~Inoue for helpful discussions. This work was supported by the JSPS (T.Y.) and by the Alfred P. Sloan Foundation (Y.T.).


\appendix

\section{Derivation of the Usadel equation}
\label{AUE}
\begin{widetext}
Here, we present the derivation of the Usadel equation in the presence of the exchange field. The matrix Green's function in particle-hole $\otimes$ spin space can be defined as\cite{Efetov1} 
\begin{eqnarray}
\hat G =  - i\left( {\begin{array}{*{20}c}
   {\left\langle {T_C a_ \uparrow  a_ \uparrow ^\dag  } \right\rangle } & {\left\langle {T_C a_ \uparrow ^{} a_ \downarrow ^\dag  } \right\rangle } & {\left\langle {T_C a_ \uparrow ^{} a_ \downarrow ^{} } \right\rangle } & {\left\langle {T_C a_ \uparrow  a_ \uparrow ^{} } \right\rangle }  \\
   {\left\langle {T_C a_ \downarrow ^{} a_ \uparrow ^\dag  } \right\rangle } & {\left\langle {T_C a_ \downarrow ^{} a_ \downarrow ^\dag  } \right\rangle } & {\left\langle {T_C a_ \downarrow ^{} a_ \downarrow ^{} } \right\rangle } & {\left\langle {T_C a_ \downarrow ^{} a_ \uparrow ^{} } \right\rangle }  \\
   {\left\langle {T_C a_ \downarrow ^\dag  a_ \uparrow ^\dag  } \right\rangle } & {\left\langle {T_C a_ \downarrow ^\dag  a_ \downarrow ^\dag  } \right\rangle } & {\left\langle {T_C a_ \downarrow ^\dag  a_ \downarrow ^{} } \right\rangle } & {\left\langle {T_C a_ \downarrow ^\dag  a_ \uparrow ^{} } \right\rangle }  \\
   {\left\langle {T_C a_ \uparrow ^\dag  a_ \uparrow ^\dag  } \right\rangle } & {\left\langle {T_C a_ \uparrow ^\dag  a_ \downarrow ^\dag  } \right\rangle } & {\left\langle {T_C a_ \uparrow ^\dag  a_ \downarrow ^{} } \right\rangle } & {\left\langle {T_C a_ \uparrow^\dag  a_ \uparrow   } \right\rangle }
\end{array}} \right)\,,
\end{eqnarray}
where $T_C$ is the time-ordering operator along the Keldysh time contour. 
This basis is obtained from the conventional basis,
\begin{eqnarray}
\hat G =  - i\left( {\begin{array}{*{20}c}
   {\left\langle {T_C a_ \uparrow  a_ \uparrow ^\dag  } \right\rangle } & {\left\langle {T_C a_ \uparrow ^{} a_ \downarrow ^\dag  } \right\rangle } & {\left\langle {T_C a_ \uparrow  a_ \uparrow ^{} } \right\rangle } & {\left\langle {T_C a_ \uparrow ^{} a_ \downarrow ^{} } \right\rangle }  \\
   {\left\langle {T_C a_ \downarrow ^{} a_ \uparrow ^\dag  } \right\rangle } & {\left\langle {T_C a_ \downarrow ^{} a_ \downarrow ^\dag  } \right\rangle } & {\left\langle {T_C a_ \downarrow ^{} a_ \uparrow ^{} } \right\rangle } & {\left\langle {T_C a_ \downarrow ^{} a_ \downarrow ^{} } \right\rangle }  \\
   {\left\langle {T_C a_ \uparrow ^\dag  a_ \uparrow ^\dag  } \right\rangle } & {\left\langle {T_C a_ \uparrow ^\dag  a_ \downarrow ^\dag  } \right\rangle } & {\left\langle {T_C a_ \uparrow ^\dag  a_ \uparrow  } \right\rangle } & {\left\langle {T_C a_ \uparrow ^\dag  a_ \downarrow ^{} } \right\rangle }  \\
   {\left\langle {T_C a_ \downarrow ^\dag  a_ \uparrow ^\dag  } \right\rangle } & {\left\langle {T_C a_ \downarrow ^\dag  a_ \downarrow ^\dag  } \right\rangle } & {\left\langle {T_C a_ \downarrow ^\dag  a_ \uparrow ^{} } \right\rangle } & {\left\langle {T_C a_ \downarrow ^\dag  a_ \downarrow ^{} } \right\rangle }
\end{array} } \right)\,,
\end{eqnarray}
by the following transformation:
\begin{eqnarray}
\hat G \to U \hat G U^\dag ,
\end{eqnarray}
with
\begin{eqnarray}
\hat U = \left( {\begin{array}{*{20}c}
   1 & 0  \\
   0 & {\sigma _1 }  \\
\end{array}} \right).
\end{eqnarray}
With the quasiclassical approximation, we obtain  the quasiclassical Green's function $\hat g$ from $\hat G$ in this basis. \cite{Efetov1} 

Then, the pair potential is transformed as 
\begin{eqnarray}
\hat \Delta  &=& \left( {\begin{array}{*{20}c}
   {0} & {\Delta \sigma _2 }  \\
   {\Delta ^* \sigma _2 } & {0}
\end{array}} \right) = (\tau_1{\rm Re}\,\Delta  - \tau_2{\rm Im}\,\Delta) \otimes \sigma _2  \\ 
  &\to& \left( {\begin{array}{*{20}c}
   {0} & {\Delta \sigma _2 \sigma _1 }  \\
   { \Delta ^* \sigma _1\sigma _2 } & {0} 
\end{array}} \right) = \left( {\begin{array}{*{20}c}
   {0} & { - i\Delta \sigma _3 }  \\
   {\Delta ^* \sigma _3 } & {0} 
\end{array}} \right) = (\tau_2\rm Re\,\Delta + \tau_1{\rm Im}\,\Delta) \otimes \sigma _3 \,,
\end{eqnarray}
and also we have 
\begin{eqnarray}
\hat{\bm{\sigma}}  = \left( {\begin{array}{*{20}c}
   \bm{\sigma}  & 0  \\
   0 & {\bm{\sigma} ^* }  \\
\end{array}} \right) \to \left( {\begin{array}{*{20}c}
   \bm{\sigma }  & 0  \\
   0 & {\sigma _1 \bm{\sigma } ^* \sigma _1 }  \\
\end{array}} \right) = (\tau _0  \otimes \sigma _1 ,\tau _0  \otimes \sigma _2 ,\tau _3  \otimes \sigma _3 ) \equiv \hat{\bm{S }}\,.
\end{eqnarray}
The Usadel equation is, therefore, transformed by $\hat U$ as\cite{Efetov1}
\begin{eqnarray}
 D\boldsymbol{\nabla} (\hat g\boldsymbol{\nabla} \hat g) +[ {i \varepsilon \tau _3  - i(\mathbf{h} \cdot \hat{\bm{\sigma}} ) - (\tau_1{\rm Re}\,\Delta  - \tau_2{\rm Im}\,\Delta) \otimes \sigma _2 ,\hat g} ] &=& 0 \\ 
  \to D\boldsymbol{\nabla} (\hat g\boldsymbol{\nabla} \hat g) + [ {i \varepsilon \tau _3  - i(\mathbf{h} \cdot \hat{\bm{S }}) - (\tau_2{\rm Re}\,\Delta + \tau_1{\rm Im}\,\Delta) \otimes \sigma _3 ,\hat g} ] &=& 0\,.
\end{eqnarray}

Next, we consider the following unitary transformation:\cite{ivanov_prb_06}
\begin{eqnarray}
V = \exp \left[ {i\frac{\pi }{4}\tau _3  \otimes \sigma _3 } \right]\exp \left[ { - i\frac{\pi }{4}\sigma _3 } \right]\,,
\end{eqnarray}
such that
\begin{eqnarray}
 V\sigma _1 V^\dag   = \sigma _1  \otimes \tau _3\,, \;
 V\sigma _2 V^\dag   = \sigma _2  \otimes \tau _3\,,  \;
 V\tau _2  \otimes \sigma _3 V^\dag   = \tau _1\,,  \;
 V\tau _1  \otimes \sigma _3 V^\dag   =  - \tau _2\,,
\end{eqnarray}
employing the relation
\begin{eqnarray}
e^{i\theta \mathbf{n} \cdot \bm{\sigma}}= \cos \theta  + i\sin \theta \mathbf{n} \cdot \bm{\sigma}\,,
\end{eqnarray}
for an arbitrary unit vector $\mathbf{n}$.
The Usadel equation is finally transformed by $V$ as\cite{ivanov_prb_06}
\begin{eqnarray}
 D\boldsymbol{\nabla} (\hat g\boldsymbol{\nabla} \hat g) +[ {i \varepsilon \tau _3  - i(\mathbf{h} \cdot \hat{\bm{S }}) - (\tau_2{\rm Re}\,\Delta + \tau_1{\rm Im}\,\Delta) \otimes \sigma _3 ,\hat g}] &=& 0 \\ 
  \to D\boldsymbol{\nabla} (\hat g\boldsymbol{\nabla} \hat g)+[ {i \varepsilon \tau _3  - i(\mathbf{h} \cdot \bm{\sigma} ) \otimes \tau _3  - (\tau_1{\rm Re}\,\Delta - \tau_2{\rm Im}\,\Delta) \otimes \sigma _0 ,\hat g}]& =& 0\,.
\end{eqnarray}
This representation of the Usadel equation for the transformed Green's function has the convenience of the explicit symmetry with respect to rotations of the exchange field $\mathbf{h}$.

Assuming that $\Delta$ is real, the equation for the $f$ component yields
\begin{eqnarray}
D\boldsymbol{\nabla} (g\boldsymbol{\nabla} f + f\boldsymbol{\nabla} \tilde g) + 2i \varepsilon f - i(f \mathbf{h} \cdot \bm{\sigma } + \mathbf{h} \cdot \bm{\sigma } f) = \Delta \tilde g - \Delta g\,.
\end{eqnarray}
Linearizing this equation with respect to superconducting correlations, we find
\begin{eqnarray}
D \boldsymbol{\nabla}^2 f + 2i \varepsilon f - i(fh \cdot \bm{\sigma }  + \mathbf{h} \cdot \bm{\sigma } f) =  - 2\Delta\,.
\end{eqnarray}
Setting $f = f_s  + \mathbf{f}_t  \cdot \bm{\sigma }$,\cite{champel} we have
\begin{eqnarray}
D \boldsymbol{\nabla}^2 (f_s  + \mathbf{f}_t  \cdot \bm{\sigma } ) + 2i \varepsilon (f_s  + \mathbf{f}_t  \cdot \bm{\sigma } ) - i(2f_s \mathbf{h} \cdot \bm{\sigma }  + 2\mathbf{f}_t  \cdot \mathbf{h}) =  - 2\Delta\,,
\end{eqnarray}
giving finally:
\begin{eqnarray}
D \boldsymbol{\nabla}^2 f_s  + 2i \varepsilon f_s  - 2i\mathbf{f}_t  \cdot \mathbf{h} &=&  - 2\Delta\,,  \\ 
D \boldsymbol{\nabla}^2 \mathbf{f}_t  + 2i \varepsilon \mathbf{f}_t  - 2if_s \mathbf{h} &=& 0 \,.
\end{eqnarray}
Here, $\mathbf{f}_t$ and $\mathbf{h}$ are three-dimensional vectors. As shown in Ref.~\onlinecite{champel}, $f_s$ and $\mathbf{f}_t \parallel \mathbf{h}$ are short ranged while $\mathbf{f}_t  \bot \mathbf{h}$ is long ranged, at low energies.
\end{widetext}

\section{Solution of the Usadel equation}
\label{SUE}

Solving Eqs.~(\ref{solution})-(\ref{bc4}), we obtain the solution of the Usadel equation as follows:
\begin{eqnarray}
C = \frac{- a_3 a_4 a_7  - a_1 a_6 a_8  + a_1 a_4 a_{10}}{a_2 a_4 a_7  + a_1 a_5 a_8  - a_1 a_4 a_9 }\,,
\end{eqnarray}
$A =  - (a_2 C + a_3 )/a_1$, $B =  - (a_5 C + a_6 )/a_4$,  $A' = b_1 A + b_2$, $B' = b_3 B + b_4$, and $C' = b_5 C$, where
\begin{align}
a_1  &= 1 - i\gamma _B \xi k_ +   - i\gamma _\phi  \cos \theta  + (1 + i\gamma _B \xi k_ +   - i\gamma _\phi  \cos \theta )b_1 \,,\nonumber \\
 a_2  &= i\gamma _\phi  (1 + b_5 )\sin \theta /2\,,\nonumber  \\
 a_3  &= (1 + i\gamma _B \xi k_ +   - i\gamma _\phi  \cos \theta )b_2 \,,\nonumber \\
 a_4  &= 1 - i\gamma _B \xi k_ -   + i\gamma _\phi  \cos \theta  + (1 + i\gamma _B \xi k_ -   + i\gamma _\phi  \cos \theta )b_3 \,,\nonumber \\
 a_5  &= i\gamma _\phi  (1 + b_5 )\sin \theta /2\,,\nonumber \\
 a_6  &= (1 + i\gamma _B \xi k_ -   + i\gamma _\phi  \cos \theta )b_4 \,,\nonumber \\
 a_7  &= i\gamma _\phi  \sin \theta (1 + b_1 )\,,\nonumber \\
 a_8  &= i\gamma _\phi  \sin \theta (1 + b_3 )\,,\nonumber \\
 a_9  &= 1 - i\gamma _B \xi k_0  + (1 + i\gamma _B \xi k_0 )b_5 \,,\nonumber \\
 a_{10}  &= i\gamma _\phi  \sin \theta (b_2  + b_4 )\,,\nonumber
\end{align}
and
\begin{align}
 b_1 &=  - \frac{{g_S^0  - g_S^3  + i\gamma _B \xi k_ +  }}{{g_S^0  - g_S^3  - i\gamma _B \xi k_ +  }}e^{2ik_ +  L} \,,\nonumber \\
 b_2  &= \frac{{e^{ik_ +  L} f_S^ -  }}{{g_S^0  - g_S^3  - i\gamma _B \xi k_ +  }}\,,\nonumber \\
 b_3  &=  - \frac{{g_S^0  + g_S^3  + i\gamma _B \xi k_ -  }}{{g_S^0  + g_S^3  - i\gamma _B \xi k_ -  }}e^{2ik_ -  L} \,,\nonumber \\ 
 b_4 & = \frac{{e^{ik_ -  L} f_S^ +  }}{{g_S^0  + g_S^3  - i\gamma _B \xi k_ -  }}\,,\nonumber \\
 b_5  &=  - \frac{{g_S^0  + i\gamma _B \xi k_0 }}{{g_S^0  - i\gamma _B \xi k_0 }}e^{2ik_0 L} \,,\nonumber
\end{align}
with $f_S^ \pm   = (f_S^0  \pm f_S^3 )/2$.

%


\end{document}